\documentclass[preprint,number,sort&compress]{elsarticle}

\usepackage{lineno,hyperref}%
\usepackage{graphicx, epsfig, amsmath, amssymb, afterpage,makeidx, setspace, subfigure}
\usepackage{color}
\usepackage[english]{babel}

\journal{Journal of \LaTeX\ Templates}

\begin{document}

\begin{frontmatter}

\title{Novel diagnostic for precise measurement of the modulation frequency of Seeded Self-Modulation via Coherent Transition Radiation in AWAKE}


\author[MPP]{F. Braunmueller}
\ead{braunmue@mpp.mpg.de}
\author[MPP]{M. Martyanov}
\author[EPFL]{S. Alberti}
\author[MPP,CERN]{P. Muggli}
\address[MPP]{Max Planck Institut for Physics Munich, F{\"o}hringer Ring 6, 80805 M{\"u}nchen, Germany}
\address[EPFL]{\'Ecole Polytechnique F\'ed\'erale de Lausanne (EPFL), Swiss Plasma Center (SPC), CH-1015 Lausanne, Switzerland}
\address[CERN]{CERN, 1022 Geneva, Switzerland}
\begin{abstract}
We present the set-up and test-measurements of a waveguide-integrated heterodyne diagnostic for coherent transition radiation (CTR) in the AWAKE experiment. The goal of the proof-of-principle experiment AWAKE is to accelerate a witness electron bunch in the plasma wakefield of a long proton bunch that is transformed by Seeded Self-Modulation (SSM) into a train of proton micro-bunches. The CTR pulse of the self-modulated proton bunch is expected to have a frequency in the range of 90-300$\,$GHz and a duration of 300-700$\,$ps. The diagnostic set-up, which is designed to precisely measure the frequency and shape of this CTR-pulse, consists of two waveguide-integrated receivers that are able to measure simultaneously. They cover a significant fraction of the available plasma frequencies: the bandwidth 90-140$\,$GHz as well as the bandwidth 255-270$\,$GHz or 170-260$\,$GHz in an earlier or a latter version of the set-up, respectively. The two mixers convert the CTR into a signal in the range of 5-20$\,$GHz that is measured on a fast oscilloscope, with a high spectral resolution of 1-3$\,$GHz dominated by the pulse length. In this contribution, we describe the measurement principle, the experimental set-up and a benchmarking of the diagnostic in AWAKE.
\end{abstract}

\begin{keyword}
AWAKE \sep Plasma Wakefield Acceleration \sep  Coherent Transition Radiation \sep  sub-TeraHertz \sep  Heterodyne Mixing
\MSC[2010] 00-01\sep  99-00
\end{keyword}

\end{frontmatter}


\section{Introduction}
\begin{figure}[!ht]
\centerline{\includegraphics[width=1.05\linewidth]{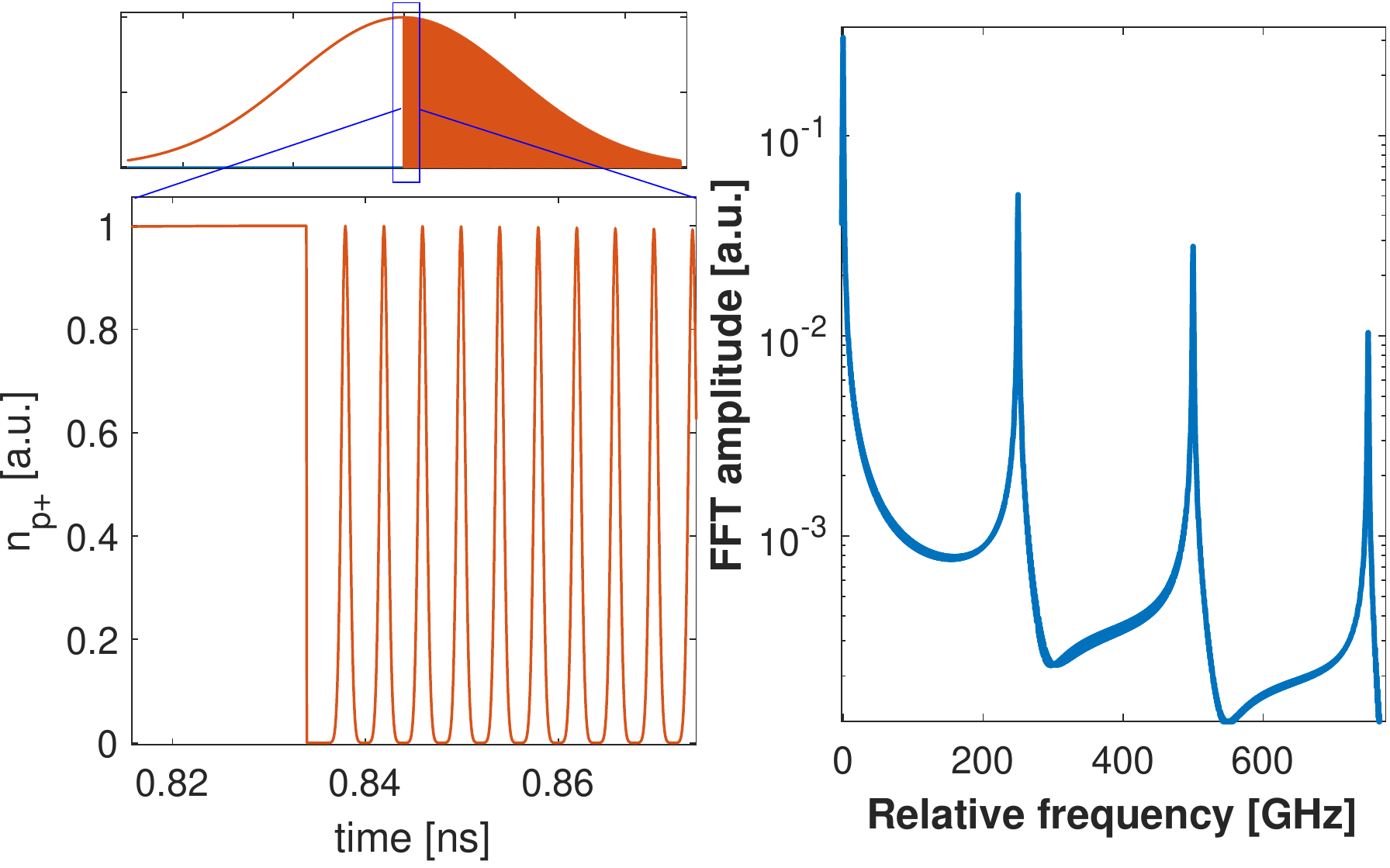}}
 \caption{\label{Fig_CTRcalc}Model of self-modulated proton bunch: central proton density close to the seeding position (left bottom), full bunch profile (left top) and FFT (right) of modeled bunch ($\sigma_z=10\,$cm). The micro-bunches in the modulated half of the bunch result in spectral peaks at the modulation frequency $f_{\textrm{pe}}=250\,$GHz and its harmonics.}
\end{figure}

\newcommand\SSM{Seeded Self Modulation }
The novel accelerator concept of Plasma Wakefield Acceleration (PWFA) consists in the acceleration of particles in the strong accelerating field (up to several GV/m) behind a relativistic particle bunch in a plasma \cite{ChenPWFA85, muggliMeterPWFA04}. AWAKE \cite{AWAKEcollabDescrPaper14, CaldwellPath2AWAKE15} is the world's first proton-driven PWFA experiment, and is located at CERN. The long proton bunches ($\sigma_z \approx 6-12\,$cm) provided by the CERN Super Proton Synchrotron (SPS) do not effectively excite wakefields. Therefore, AWAKE relies on the seeded self-modulation (SSM) of the proton bunch \cite{KumarFirstSMI10, LotovSeededSMI13}, turning the proton bunch into a series of micro-bunches with a spacing of the plasma wavelength, which are well-suited for exciting large amplitude wakefields. The seeding of the self-modulation is provided by the sharp ionization front created by a 4.5$\,$TW / 120$\,$fs laser pulse. The laser pulse is co-propagating within the proton bunch, singly ionizing the Rubidium (Rb) vapour to form a channel of uniform plasma $\sim\!1\,$mm radius over the 10$\,$m vapour source. The SSM is a transverse process that periodically focuses and defocuses the protons, therefore locally increasing/decreasing the proton density with a longitudinal periodicity. Along the bunch axis, a periodic train of short micro-bunches persists, as shown in a series of publications \cite{MuggliAwakePhysicsIPAC13,  LotovphysSMIPoP15, VieiraSMIFinRise14}. A witness electron bunch will be injected in the wakefields of the self-modulated proton bunch for acceleration experiments. \\
The role of the diagnostic described here is to precisely measure the frequency emitted by the self-modulated bunch. The diagnostic measures the coherent transition radiation (CTR) of the self-modulated proton bunch passing through a metallic screen. Each proton passing through the screen creates transition radiation at all frequencies, which adds up coherently for wavelengths longer than the the typical longitudinal and radial structure of the bunch. \\
 The CTR radiation pattern and energy have been calculated for self-modulated bunches and will be published elsewhere \cite{MishaCTR}. These calculations show that the CTR signal at the modulation frequency occurs and should be strong enough to be detected in a single shot by standard microwave diagnostics such as the one described here.\\
For illustrating the CTR frequency-content of the self-modulated bunch, the bunch density close to its central axis is modelled in Fig. \ref{Fig_CTRcalc} by a gaussian bunch envelope with $\sigma_z=10\,$cm (see left top Fig. \ref{Fig_CTRcalc}) and a series of gaussian micro-bunches with $\sigma_z=0.1\,$mm and a modulation frequency of 250$\,$GHz in the second half of the bunch envelope (zoom in left bottom Fig. \ref{Fig_CTRcalc}). 
This simple model does not realistically represent the shape of the micro-bunches, but it can be used to show how the shape and periodicity of the microbunches translates into the shape of the FFT-peaks. Simple physics as in this model says that the radiation spectrum of the train of microbunches has a spectrum with peaks at the modulation frequency and its harmonics, as shown in the FFT of the modelled bunch density on the right side of Fig.\ref{Fig_CTRcalc}.\\
The modulation frequency is predicted to be very close to the plasma frequency \cite{LotovphysSMIPoP15} $f_{\textrm{pe}} \sim \sqrt{n_e}$, depending on the plasma electron density $n_e$. The plasma frequencies in AWAKE are in the range of 90-300$\,$GHz ($n_e=10^{14}-10^{15}$cm$^{-3}$). The diagnostic described here will attempt to show that the CTR-frequency, and therefore the bunch modulation frequency, corresponds to this plasma frequency, as predicted by the SSM models.
\section{\label{sec_measPrinciple} Measurement principle}
Signals in the range of $f_{\textrm{RF}}=$90-300$\,$GHz can generally not be sampled directly and are therefore measured either with a spectral method or by a heterodyne mixing method. \\
The principle of heterodyne mixing \cite{brundermann2012THztechn,bauerMMwaveMixer66}, which is employed in our case for the detection of sub-Terahertz radiation, relies on the detection of an intermediate beat frequency ($f_{\textrm{IF}}$) between the unknown-frequency RF-signal ($f_{\textrm{RF}}$) and a nearby reference frequency ($f_{\textrm{ref}}$): $f_{\textrm{IF}} = \left\rvert f_{\textrm{RF}} - f_{\textrm{ref}} \right\rvert
$. Here, the modulus indicates that the mixing can result from two 'branches', with $ f_{\textrm{RF}} > f_{\textrm{ref}}$ or $ f_{\textrm{RF}} < f_{\textrm{ref}}$. \\
For a direct measurement of the IF-signal, a low frequency 5$\,$GHz$\leq f_{\textrm{IF}} < 25\,$GHz is preferable, so that the reference frequency has to be close to the RF-frequency $f_{\textrm{RF}} \approx f_{\textrm{ref}}$. The tunable reference frequency is therefore generated by a frequency-multiplication of a lower-frequency local-oscillator ($f_{\textrm{LO}}$): 
$f_{\textrm{ref}}  =  n_{\textrm{harm}} \cdot f_{\textrm{LO}}$ with 10$\,$GHz $\lesssim f_{\textrm{LO}} \lesssim 23\,$GHz and harmonic number $n_{\textrm{harm}} \in [8,..,30]$. 
In the frequency-multiplication of $f_{\textrm{LO}}$, several harmonics are created. Since the $f_{\textrm{IF}}$ signal can originate from various harmonics of $f_{\textrm{LO}}$, the harmonic number $n_{\textrm{harm}}$ must be determined for each frequency measurement. This is done by plotting $f_{\textrm{IF}}$ as a function of $f_{\textrm{LO}}$ for a constant $f_{\textrm{RF}}$: $n_{\textrm{harm}} = \frac{\Delta f_{\textrm{IF}}}{\Delta f_{\textrm{LO}}}$. 
%

\section{\label{sec_ExpSetup}Experimental set-up}
\begin{figure}[!ht]
 \centerline{\includegraphics[width=0.65\linewidth]{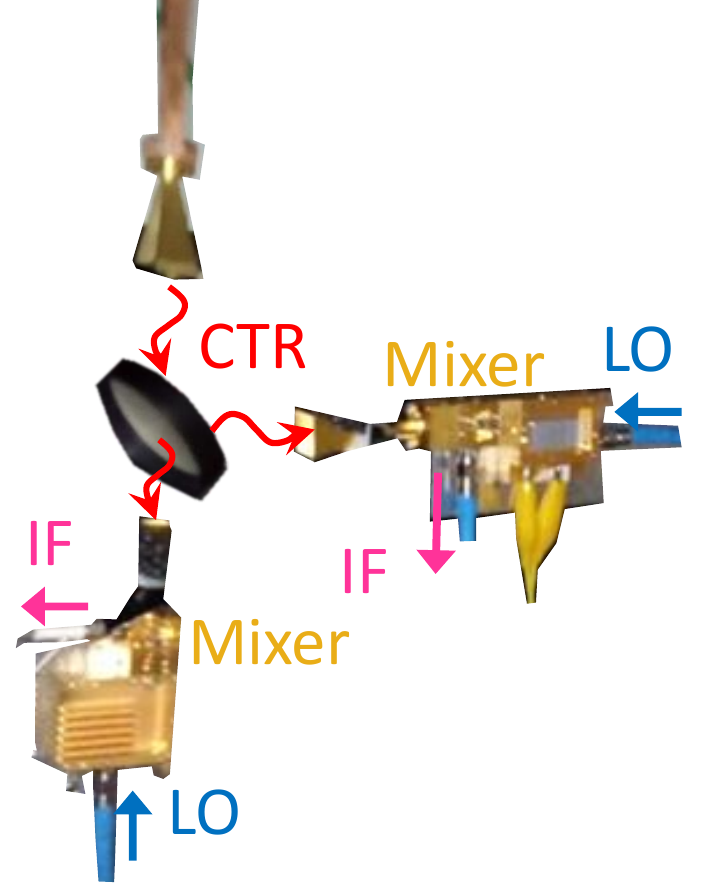}}
 \caption{\label{fig_PicDetectors}Sketch/Photo-excerpts of the CTR-detector set-up. The CTR-signal, coupled out of the transmission line by a horn-antenna, is divided by a beam splitter between a lower frequency-band heterodyne mixer (WR8 / 90-140$\,$GHz) and a higher frequency-band mixer (WR3.4 / 255-270$\,$GHz or WR4.3 / 170-260$\,$GHz).}
\end{figure}

\begin{figure}[!ht]
 \centerline{\includegraphics[width=1.0\linewidth]{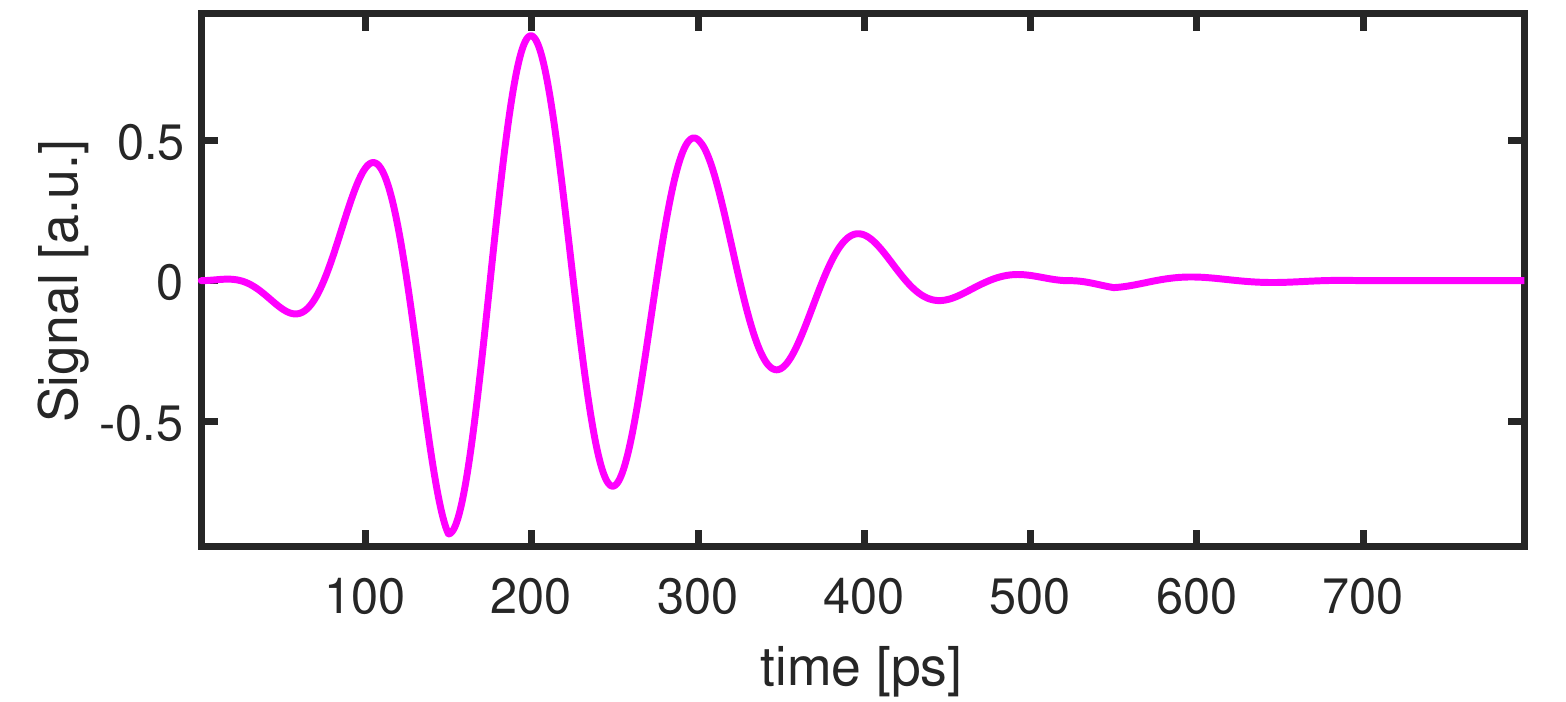}}
 \caption{\label{fig_FakeIFsignal}Expected IF-signal for $f_{\text{IF}} = 10\,\text{GHz} $.}
\end{figure}

The CTR is emitted on an Al-coated SiO screen with radial polarization, in a cone of $\sim$5-20$^{\circ}$ half angle, and with a peak power of $\sim 10-30\,$kW/sr \cite{MishaCTR}. The CTR is then collimated by a 45$^{\circ}$ / 4" off-axis parabola (OAP) mirror. Various diagnostics are placed near the beamline and at the end of the CTR transmission line, of which we describe only the waveguide-based heterodyne detectors. \\
Because this diagnostic is placed behind a shielding wall, a transport of the CTR-signal over 15 meters is required between the CTR-screen in the beamline and the detector. A flanged WR90 rectangular waveguide (8-12$\,$GHz) is placed without any coupler in the collimated beam and collects a small fraction, estimated to be $\sim 2-3\,$\%, of the CTR signal (90-300$\,$GHz).\\
In order to minimise the losses, the transport over a 14$\,$m waveguide section is carried out in the WR90 waveguide, using the TE$_{01}$ 'tall' waveguide mode, and is therefore very overmoded.
The losses and mode conversion in the transmission line are caused by Ohmic losses and small unavoidable tilts or bends in the straight sections, as well as by mode conversion in the four 90$^{\circ}$ mitered waveguide corners and in the final waveguide taper. From calculations of each of these losses, the overall transmission efficiency of the waveguide is estimated to be of the order of one quarter.
\\
At the end of the transmission line, a 39$\,$cm non-linear waveguide taper provides a smooth transition to the fundamental TE$_{01}$-mode of a WR8 pyramidal horn antenna. This tapered waveguide simultaneously rejects unwanted higher-order waveguide modes and acts as a high-pass filter with a cutoff-frequency of 74$\,$GHz. This cut-off rejects the CTR frequencies from the unmodulated bunch envelope ($f_{CTR} \gtrsim 20\,$GHz, blocking the low-frequency peak in Fig. \ref{Fig_CTRcalc}, right side). The waveguide-based heterodyne mixers, collect the CTR-signal via another pyramidal horn antenna. \\
The heterodyne detectors consist of the frequency-tunable LO-synthesizer, an amplifier/frequency multiplier chain (AMC), and a sub-harmonic mixer. In the mixer, the RF-signal is mixed with twice the input reference frequency by a mixer based on a pair of anti-parallel diodes. Furthermore, unwanted LO-harmonics can be suppressed in the AMC by a frequency-filter, and an attenuator can be used for adapting the power of the reference signal into the mixer.\\
Three different heterodyne systems of this type have been used, as listed in table \ref{table_SignGens}, with varying configurations in different experimental campaigns. These detectors have been purchased as a complete set from the manufacturers, including off-the-shelf mixers (Radiometer Physics\textsuperscript{\textregistered} SHM 90-140 \& SHM 170-260 and Virginia Diodes\textsuperscript{\textregistered} WR3.4SHM-S009c) .
\begin{table*}[ht]
\centering
\begin{spacing}{1.0}
\begin{tabular}{|c|c|c|c|c|}
\hline
Frequency & Waveguide & $ $ & LO-harmonics & LO-frequency  \\
range & type & Manufacturer & $n_{\text{harm}}$ & $f_{\text{LO}}$   \\
\hline
255-270$\,$GHz & WR3.4 & Virginia Diodes Inc\textsuperscript{\textregistered} (VDI) & 24 & 10-12$\,$GHz \\
90-140$\,$GHz & WR8 & Radiometer Physics\textsuperscript{\textregistered} & 8 & 11.25-17.5$\,$GHz \\
170-260$\,$GHz & WR4.3 & Radiometer Physics\textsuperscript{\textregistered}  & 12 & 14.17-21.67$\,$GHz \\
\hline
\end{tabular}
\end{spacing}%
\caption{\label{table_SignGens} Specifications of different waveguide-integrated heterodyne CTR-detectors used in AWAKE, in chronological order of application in AWAKE.}
\end{table*}
%
The WR3-detector, which is on loan from the Swiss Plasma Center at the EPFL university, has an on-board local oscillator, while the WR4 and WR8 detectors are supplied by external tunable synthesizers (Hewlett-Packard\textsuperscript{\textregistered} 83642A and Vaunix \textsuperscript{\textregistered} LMS-203). \\
The important components of the detector set-up are shown in Figure \ref{fig_PicDetectors}. 
%
The CTR-signal coming from the common transmission line is divided by a flip-mounted beam splitter between a 90-140$\,$GHz detector and either a 255-270$\,$GHz mixer or a 170-260$\,$GHz mixer. The beam splitter, used in p-polarization, consists of a 1$\,$mm-thickness HRFZ-Si wafer (High Reflectivity Flat Zone Si), with a frequency-insensitive 46/54 splitting ratio \cite{tydexBSWebSite}. This allows for covering a large fraction of the possible modulation frequencies. For a certain low plasma density and frequency-range, the set-up also allows for measuring the second harmonic of the modulation-frequency in the CTR-signal at the same time as its fundamental frequency. Higher harmonics content is an indication for a deep SSM producing microbunches shorter than the modulation period. \\
The IF-signal created by the mixer is amplified in a 0-24$\,$GHz low-noise amplifier and coupled via low-loss SMA-cables into a fast oscilloscope with 23$\,$GHz-bandwidth (Tektronix\textsuperscript{\textregistered} MSO72304DX), where the alternating voltage of the IF-signal is directly acquired. \\
The length of the CTR-signal is expected to be as short as the modulated part of the proton bunch, which is of the order of 300-700$\,$ps. The expected IF-signal from this bunch is illustrated in Figure \ref{fig_FakeIFsignal}, showing a calculated signal with $f_{\text{IF}} =10\,$GHz and a width of 300$\,$ps (FWHM). 
This illustrates that the intermediate frequency is restricted to $f_{\text{IF}} \gtrsim 5\,\text{GHz} $ in order to have several oscillation periods in the IF-signal.
\section{Signal processing and benchmarking}
\begin{figure}[!ht]
\centerline{\includegraphics[width=1.0\linewidth]{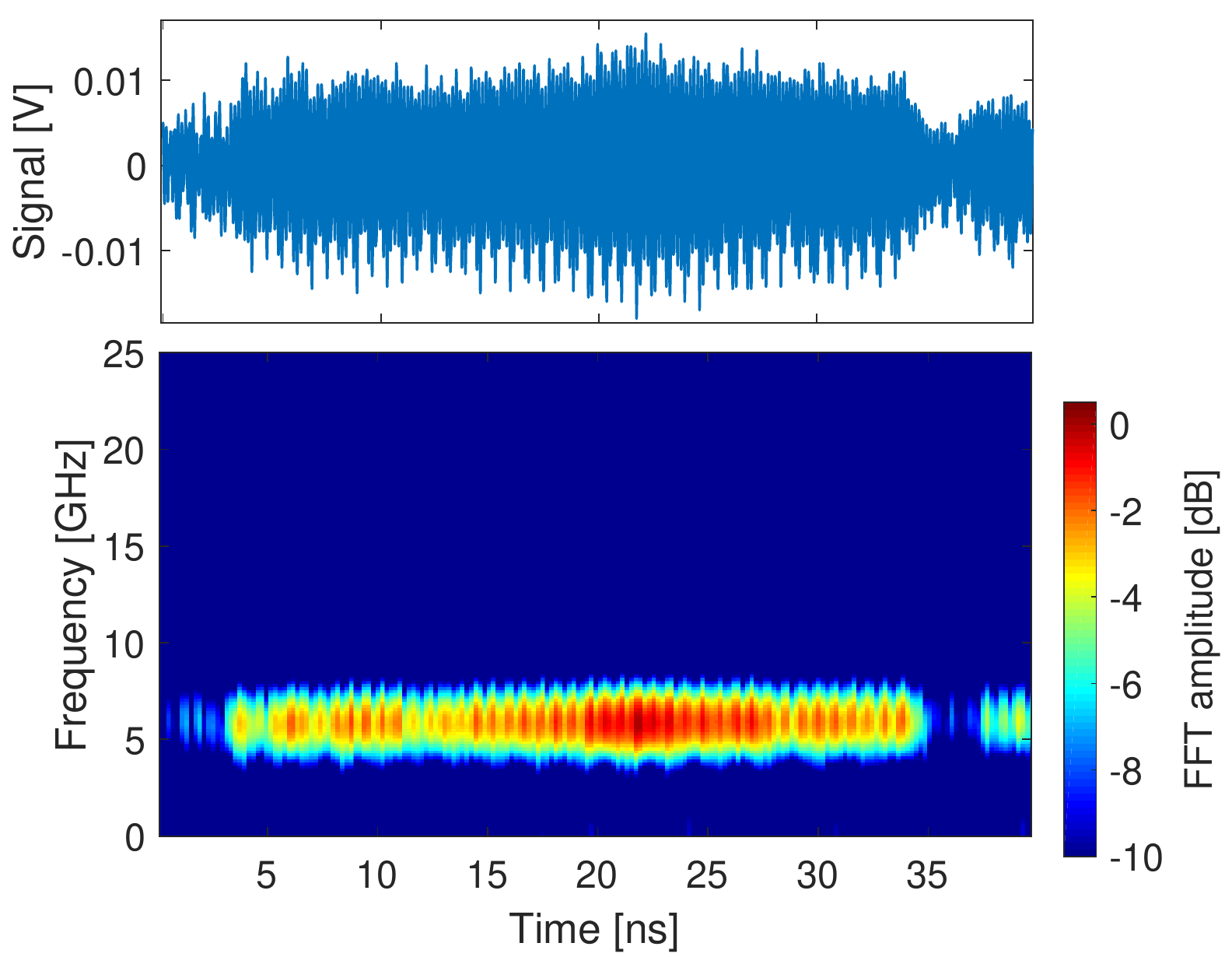}}
 \caption{\label{fig_IFMishasSource}Raw oscilloscope time-trace  (top)and spectrogram (bottom) of IF-signal from a tes	t-measurement of a pulsed RF-source with $f_{\text{RF}}=110.1$\,GHz.}
\end{figure}

\begin{figure}[!hb]
\centerline{\includegraphics[width=1.0\linewidth]{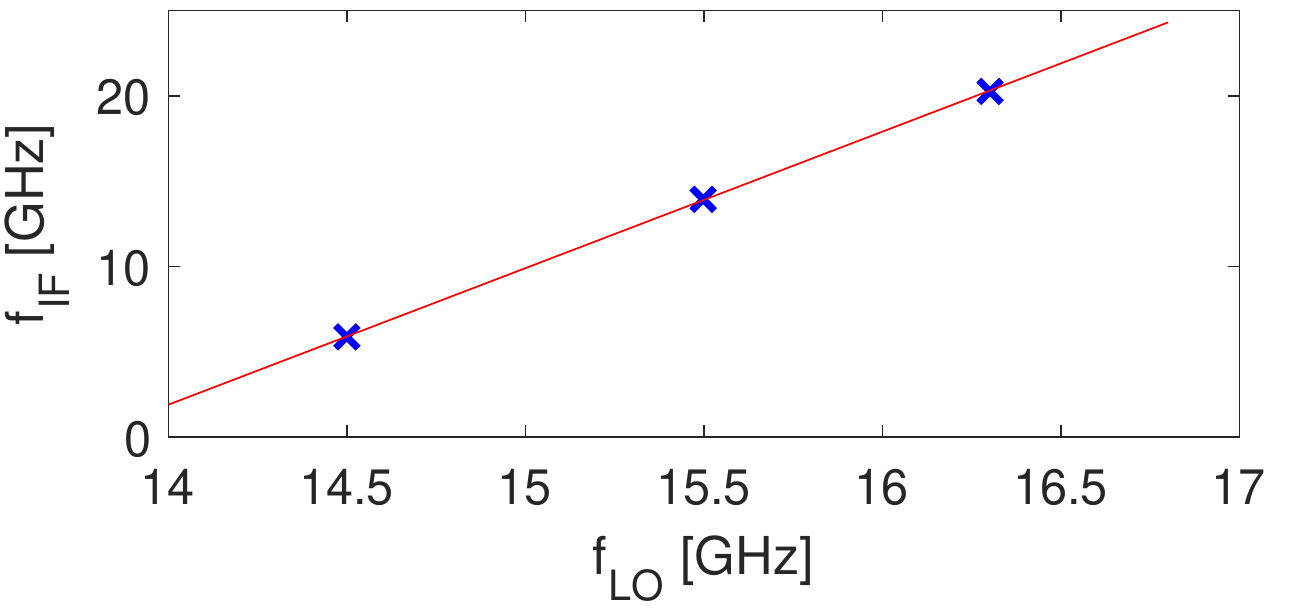}}
 \caption{\label{fig_LOscanMishasSource}IF-frequency of test-measurements of RF-signal from fixed-frequency source at 110.1$\,$GHz as in Fig. \ref{fig_IFMishasSource}, determined via FFT (longer windowing), as a function of local-oscillator frequency. The fit yields $n_{\text{harm}}=8.00$.}
\end{figure}

The 300-700$\,$ps signal is analysed in a spectrogram (spectrum versus time) using a windowing (zero-padded) FFT, in which consecutive 520$\,$ps FFT-windows are shifted by 260$\,$ps. The IF-frequency is taken from the time with the highest FFT-amplitude in the 5-22$\,$GHz-bandwidth. 
\smallbreak
Frequency uncertainties in the heterodyne detection can result from uncertainties in the LO-frequency, from uncertainties in the sampling rate of the oscilloscope, or from uncertainties in the frequency analysis of the signal. These three components have been tested by precisely measuring the frequency of the local-oscillator on the fast oscilloscope, using the frequency-detection algorithm that will be used on the CTR-signal.
This test was performed for the two local oscillators (HP and Vaunix). It was shown, that the $f_{\text{LO}}$ measured on the oscilloscope corresponds to the set $f_{\text{LO}}$ to within $< 0.1\,$GHz. This gives an estimation of the uncertainties of $f_{\text{LO}}$, the oscilloscope and the frequency detection algorithm combined.\\
Finally, a more realistic test of the detector was performed by measuring the signal of a pulsed RF-source with $f_{\textrm{RF}}\approx110\,$GHz. This source was placed at the beginning of the transmission line, with the WR8-detector (90-140$\,$GHz) measuring at the end of the transmission line. The IF-signal of the pulse is shown in Fig. \ref{fig_IFMishasSource}, both as a raw time-trace (top) and in a spectrogram (bottom).
The signal was analysed with a spectrogram using a 520$\,$ps windowing, zero-padded FFT, as used for the short-pulse experimental data. Due to the short duration of the CTR-signal (300-700$\,$ps), its signal bandwidth is 1-3$\,$GHz. The short signal is mimicked here by the short windowing, resulting in the 1-2$\,$GHz width of the frequency-peak determined from the spectrogram of Fig. \ref{fig_IFMishasSource}. The maximum of the peak is $f_{\text{IF}}=5.9$\,GHz, with the local oscillator being set to $f_{\text{RF}}=14.5$\,GHz.\\
The RF-signal is measured for different LO-frequency settings, in order to determine whether the CTR-frequency is above or below the reference frequency and to determine the LO-harmonic $n_{\text{harm}}$. The result of such a set of measurements is shown in Fig. \ref{fig_LOscanMishasSource}, showing  $f_{\text{IF}}$ for three different values of $f_{\text{LO}}$, determined from FFTs with longer windowing. 
As shown in the plot, the fit confirms $n_{\text{harm}}=8$, which is the design harmonic for the used WR8-detector. The positive slope corresponds to $f_{\text{ref}} > f_{\text{CTR}}$, so that the resulting CTR-frequency from the fit is obtained as:  $f_{\text{RF}} = n_{\text{harm}} \cdot f_{\text{LO}} - f_{\text{IF}} = 8 \cdot 14.5\,\text{GHz} - 5.9\,\text{GHz} = 110.1\,\text{GHz}$. Because the frequency of the RF-source is known to be in the range $f_{\text{RF}}=110 \pm 1 \,$GHz, this measurement is accurate to within $\lesssim 1\,$GHz.\\

This shows that the limitation in the CTR-measurement is mainly given by the bandwidth of 1-3$\,$GHz from the short CTR-signal. Therefore, the diagnostic is well-designed for the detection of a CTR-signal at the expected modulation frequencies. Assuming $f_{\text{CTR}} = f_{\text{modulation}} = f_{\text{pl}} \sim \sqrt{n_e}$, measuring the CTR-frequency to within 1-2\% would allow to determine the plasma density to an accuracy of the order of 0.5-1\%.\\
First successful measurements of CTR-pulses in AWAKE using this detector have been acquired in 2017 and will be presented in a future publication.
%
\section{\label{sec_con} Summary and Conclusion}
This contribution describes in detail the set-up and measurement principle of the waveguide-based heterodyne CTR-detectors used in the AWAKE experiment. The goal of the diagnostic is to precisely measure the modulation frequency in the CTR-signal of a proton bunch that has been modulated by the SSM-mechanism. The principle of the measurement is to mix the unknown frequency with a known reference and to measure the frequency difference of $f_{\text{IF}}= |f_{\text{CTR}} - f_{\text{ref}}| \approx 5-22\,$GHz on a large bandwidth oscilloscope. The IF-frequency of a single p$^+$-bunch can then be determined using a spectrogram method, which is repeated at different LO-frequencies in order to check the sign of the frequency-difference and to confirm the correct harmonic of the local oscillator $n_{\text{harm}}$. A simple test-measurement showed the proper working of the detectors, with a frequency-resolution of few GHz that is dominated by the bandwidth of the short CTR-pulse length.
\section*{\label{sec_ackn} Acknowledgements}
Work with the WR3-detector supported by Requip, Sinergia and (No: 200020-120503/1), grants of the Swiss National Science Foundation and by the Ecole Polytechnique F\'ederale de Lausanne (EPFL).
%
%
%
%
%
%
%

\section*{References}

\bibliography{CTRhetProceedings}

\end{document}